# MetaQuestion: A web application for expert knowledge elicitation addressing plant health and applied plant ecology

Robert Fontan[1,2,3] | Christopher M. Perez[1,2,3,4] | Ashish Adhikari[1,2,3] | Romaric A. Mouafo-Tchinda[1,2,3,5] | Aaron I. Plex Sulá[1,2,3] | Jacobo Robledo[1,2,3] | Berea A. Etherton[1,2,3,6] | Manoj Choudhary[1,2,3,7] | Muhammad Aqeel Sarwar[8] | Zunaira Afzal Naveed[8,9] | Karen A. Garrett[1,2,3]

[1]Plant Pathology Department, Institute of Food and Agricultural Sciences, University of Florida, Gainesville, FL, USA
[2]Global Food Systems Institute, University of Florida, Gainesville, FL, USA
[3]Emerging Pathogens Institute, University of Florida, Gainesville, FL, USA
[4]Department of Statistics, Florida State University, Tallahassee, FL, USA
[5]Biopterre - Bioproducts Development Center, La Pocatière, Canada
[6]Santa Fe College, Gainesville, FL, USA
[7]ICAR- National Research Institute for Integrated Pest Management, New Delhi, India
[8]Center for Advanced Studies in Agriculture and Food Security, University of Agriculture Faisalabad, Pakistan
[9]Center of Agricultural Biochemistry and Biotechnology, University of Agriculture Faisalabad, Pakistan

Correspondence
R. Fontan
Email: robertfontan@gmail.com
K. A. Garrett
Email: karengarrett@ufl.edu

**Abstract**

1. Expert knowledge elicitation provides information to characterize ecological systems and management options. Linking expert knowledge elicitation with a curated question catalog supports a community of practice for ongoing improvement of question quality.

2. The MetaQuestion web app we introduce here draws on the PlantQuest catalog of questions addressing applied plant ecology and management options, making the catalog available in a flexible form for organizers of expert knowledge elicitation. Organizers can select among questions in the catalog, modify them as needed, and generate an instrument customized to their elicitation project. MetaQuestion makes available PlantQuest questions specialized for the study of invasive species such as pathogens and arthropod pests, such as geographic analyses of prevalence and network analysis of the movement of plant materials.

3. Experts answer questions in the customized instrument and their responses are compiled. For settings where internet access may be sporadic, there are options to download the instrument for experts' work and then upload responses later. MetaQuestion provides the resulting dataset in a CSV file for analysis in users' choice of software

4. Development of the PlantQuest catalog and the MetaQuestion app is ongoing, incorporating lessons learned from applications of the app. The MetaQuestion app could also be adapted to address questions from other subject areas.

# 1 | INTRODUCTION

Effective decision-making in applied ecology often depends on efficiently integrating expert judgements with objective data, such as from field experiments. Domain experts have extensive knowledge of many aspects of ecological systems, which is not readily accessible in digital information systems. Expert knowledge elicitation provides a systematic, structured, and reproducible approach for accessing and transforming expert knowledge into usable data (Morgan 2014; O'Hagan 2019; Robledo et al. 2025). For example, expert elicitation has been widely used to prioritize the conservation of endangered species as well as facilitating rapid responses to invasive species such as pathogens and pests (Dicks et al. 2021; Martin et al. 2012; McBride et al. 2012).

Several software programs analyze data generated from structured expert elicitation (e.g., the Elicitator tool, the SHELF package (https://shelf.sites.sheffield.ac.uk/), and the Excalibur package). However, there are few open-source tools for conducting expert knowledge elicitation tailored to the needs of applied ecologists. The IDEAcology interface helps prepare and implement a quantitative expert elicitation based on the IDEA protocol (Courtney Jones et al. 2023; Hemming et al. 2018). There also remain many opportunities to make expert elicitation more efficient. For example, a curated catalog of previously applied and tested questions can facilitate the application of expert elicitation instruments for a set of topics, and can be used as a knowledge repository.

Here, we introduce MetaQuestion, an open-source web application developed to streamline expert knowledge elicitation for projects studying and supporting applied plant ecology. MetaQuestion is a "meta tool" in the sense that it is a tool allowing organizers to efficiently create elicitation instruments in a hierarchical structure. The application provides organizers with a question bank of previously applied and evaluated questions and allows rapid tailoring of questions for specific targets (e.g., species, region, pathogens, and/or arthropod pests). Organizers can then provide the resulting customized elicitation instrument to experts in their project. Once experts have provided their input, MetaQuestion allows organizers to download the machine-readable, expert-elicited data, which is ready for analysis.

A new feature of MetaQuestion, in contrast to platforms such as Google Forms and Qualtrics, is the use of a curated catalog of questions that have been tested and improved for a specific field. In this case, we incorporate the PlantQuest catalog of questions related to plant health (PlantQuest Team, in preparation). The question catalog feature allows organizers to quickly develop a comprehensive instrument to capture expert opinion. To address specific aspects of plant health, custom question types are incorporated that are not readily available in most other apps, such as map-based questions to collect spatial inputs and questions about networks that generate adjacency matrices. Another feature is its local support, allowing expert participants to download the instrument and then upload their individual or group responses. MetaQuestion is designed to work both online and offline.

This note introduces MetaQuestion and its functionality, provides examples, and discusses next steps in the MetaQuestion project. **Note that the version of MetaQuestion associated with this preprint version incorporates a preliminary version of the PlantQuest question catalog; a version of MetaQuestion available in the near future, announced in an update to this preprint, will incorporate the upcoming PlantQuest version 1.0.**

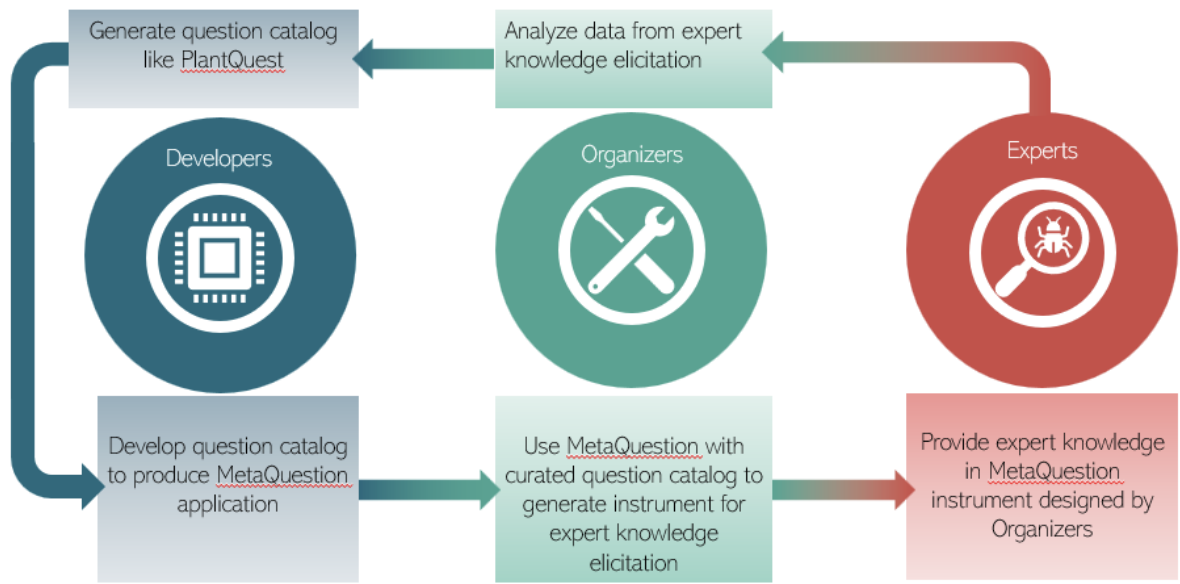

**FIGURE 1** The MetaQuestion web app draws on a curated question catalog, in this case PlantQuest, to provide a tool for organizers of expert knowledge elicitation. Organizers use MetaQuestion to generate an instrument for experts. Experts provide their knowledge. Organizers then analyze the resulting data and provide feedback for improving MetaQuestion and the question catalog.

## 2 | METAQUESTION APP

### 2.1 MAIN FEATURES

MetaQuestion is developed for two groups: 1) the organizers who prepare the instrument and questionnaire, and 2) the participating experts who answer those questions (Fig. 1). The opening page is a dashboard that allows organizers to select questions from the PlantQuest catalog. Organizers can adapt these questions or add new research questions to fit their specific needs. The "instrument constructor" page allows organizers to build out and preview their instrument, showing the various questionnaire sections, and allowing organizers to pick from the existing question sets and create new ones. Other features of MetaQuestion allow users to preview the full instrument in either a web browser or a print view, view expert answers in a table, and view simple charts, like bar graphs and pie charts, of the expert responses. Results can be downloaded in CSV format for further data analysis. (In an upcoming version of this preprint, we will provide vignettes to illustrate how to analyze CSV files in R (R Core Team 2025).) When the organizer is finished editing, the published instrument link can be provided and opened on expert participants' web browsers.

Another MetaQuestion feature is downloadable and printable versions of the instrument, so that experts in locations with inconsistent internet or power can still participate in the expert knowledge elicitation process, working offline as needed and then syncing their results when connectivity is restored. For further accessibility, MetaQuestion instruments can be exported to Qualtrics.

A range of question formats are available, designed to support topics in applied plant ecology. Basic question types include short answer responses, multiple choice, multi-select, slider, and matrix questions (including tables for responses). Unique question types, like a "map type," allow an organizer to upload an image file to act as a map and place nodes on top of it, with expert participants indicating links between labeled nodes/locations. Map type questions can be used in estimating, for example, seed movement or trade between two regions. In each question format there are options to make the question required or to change its behavior. More question types and behaviors will be incorporated, as a focus for the next stages of app development.

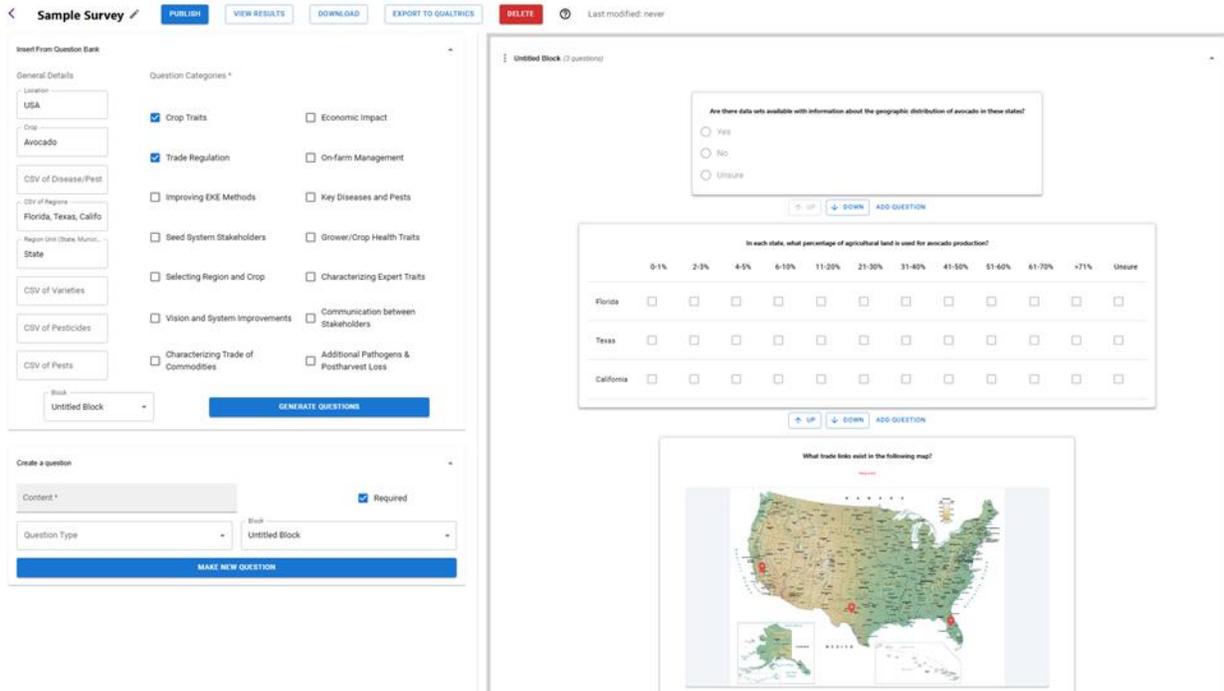

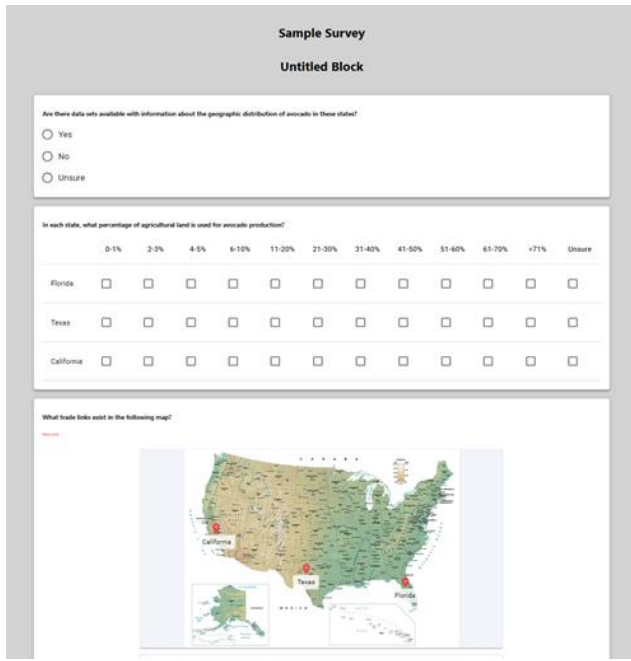

**FIGURE 2** Screenshot of editing options for organizers of expert knowledge elicitation, and illustration of a resulting instrument viewed by experts participating in the elicitation.

**FIGURE 3** Example of results table generated from expert input, as viewed by organizers of expert knowledge elicitation.

## 2.2 BASIC IMPLEMENTATION

MetaQuestion is accessed through a web browser at Netlify (https://r2meke.netlify.app/), an online platform to build and launch apps. **Note that the version of MetaQuestion associated with this preprint version incorporates a preliminary version of the PlantQuest question catalog; a version of MetaQuestion available in the near future, announced in an update to this preprint, will incorporate the upcoming PlantQuest version 1.0.** At this link, organizers create and edit an instrument for their specific project. The group of experts, assembled by the organizers, complete the questions in the instrument and the data are returned to the organizers without installing additional software. The backend of MetaQuestion is managed by SupaBase, a hosted platform built on PostgreSQL (version 14), which handles authentication, data persistence, and storage. For developers, the app and its dependencies can be installed locally; this requires Node.js (20 LTS) and Vite, with SupaBase either connected to a hosted instance or run locally via Docker. An open repository link will soon be made available at https://github.com/GarrettLab/EKE.

MetaQuestion is built using React (version 18) as a primary development framework with Vite (version 5) as the build framework. The frontend is written in JavaScript/JSX and styled with Material UI (v5) components and custom CSS. Client-side navigation is managed with React Router, enabling single-page application behavior across routes such as instrument editor, question catalog, instrument execution, and results. Pages are rendered entirely in the browser. The build pipeline uses Vite to bundle and optimize assets, producing static deployment.

## 2.3 ARCHITECTURE

**Dashboard**. The dashboard is the first page that the organizer sees when they open the application. It has a list of all the projects that the organizer has worked on previously. The organizer can select an old project to work on or create a new one. There are buttons to view the PlantQuest question bank or load a local project.

**Instrument Constructor**. The instrument constructor is where the organizer can create a new instrument. Here, the organizer can add questions, preview, and publish the instrument for use by experts. The organizer can also view the results of the expert knowledge elicitation and download the resulting data in CSV format. There is an option to download survey data in JSON format locally to the expert participants' devices to send to the participating experts, in case of any unexpected interruptions, such as sporadic internet access.

**Question Catalog**. The question catalog allows the organizer to see all the questions that have been stored from PlantQuest. The questions are organized into over a dozen distinct categories, such as crop traits, trade regulation, and economic impact.  The curated questions can also be freely edited by the organizers to adapt to their setting. The organizer can select questions to add to the instrument in the Instrument Constructor based on their specifications.

There is also the option for users to generate their own questions from the ground up. Various question types are supported, such as multiple choice, matrices, maps, short answer, and more.

**Instrument Page**. The instrument page is where expert participants answer questions organizers have incorporated. For both organizers and participants, there is a feature to generate a pdf formatted version of the instrument. A local instrument page is also available to load a JSON file in the machine in case of sporadic access to the internet.

## 2.4 OUTPUT

**Viewing Results and Output.** Once results have been collected, users can perform basic visualization of the results. For supported question types, pie and bar charts can be created using the app to visualize the data collected. Additional visualization types and support for more question types are in development for future versions of the application.

Alternatively, organizers may download the results as a CSV file. These files can be readily loaded into statistical frameworks of choice, such as R, Python, or other statistical software, allowing for integration in other analysis pipelines. Vignettes for reading the output in R and performing basic analysis will be available in a future version of this preprint.

## 2.5 DATA STORAGE & PRIVACY

Data storage is implemented using SupaBase which provides a PostgreSQL database for persistent storage. There are row-level security and authentication integration features to protect participant data. There are tables for instruments, question banks, and responses.

# 3 | TESTING

## 3.1 Code Testing

Code testing employed static analysis testing using Semgrep to detect security vulnerabilities, syntax errors, and deviations for web application codebases. Semgrep enables the scanning of React components and server-side logic using rule-based patterns. The scans were integrated into the development workflow to ensure issues were identified early. The process allows for high code quality, enforces consistency, and reduces runtime bugs or security flaws.
In addition to static analysis, unit tests were implemented for each build version of the application. These tests ensure core functionality at the component level, checking that updates did not introduce regressions and that the system behaved as expected across versions.

## 3.2 User Interface Testing

Initial user interface testing was performed in feedback sessions with members of the PlantQuest Team, who represented typical end users of the system. During the sessions, participants interacted with the application in realistic use scenarios, and their observations were documented. Feedback focused on usability, navigation, and alignment with research objectives. Next, the app was evaluated in test cases for rice health and cotton health. Organizers found the app straightforward to navigate, streamlining data collection. The app features options such as populating questions with a list of locations or species supplied by the organizers.

Prior to launching expert knowledge elicitation, the organizers compiled information about the species being evaluated from published research, official reports, and discussions with scientists who have practical experience. They then drew on the existing question catalog from PlantQuest incorporated in the app. The question catalog provides a tested framework for topics such as pest and pathogen occurrence, and movement of plant materials and related information. For example, for cotton, template questions from the catalog were edited to reflect the actual set of pathogens and arthropod pests that affect the cotton crop in Pakistan, and further adjustments were made to ensure that regional variation was captured. So, during expert knowledge elicitation to understand cotton health in Pakistan, this preparation meant that experts only saw questions that are directly relevant to cotton in Pakistan. After compiling and editing all the questions, the final instrument was published, and the link shared with the participants on the day of the expert knowledge elicitation.

The app records information about networks of species movement and connectivity through trade in plant materials. The app also supports adding open-ended questions in the instrument, where participants can provide context that may not be captured otherwise, such as observations of emerging diseases or pests, management practices that influence pathogen or pest spread, or concerns about future risks. These qualitative responses complement quantitative data and provide a fuller picture of plant health.

Current hosting of the app successfully handled the online traffic of 30 simultaneous expert users. From the organizers' perspective, the app transforms what was previously a labor-intensive and sometimes cumbersome task into a concise, digital workflow. The data gathered feeds directly into analyses without the need for lengthy post-processing. In this way, the app supports rapid and reliable assessment of plant health risks, while making efficient use of expert time and knowledge.

The results from app use help organizers, including stakeholders and policy makers, spot vulnerable species or geographic locations, and evaluate whether the management strategies growers use are working. For researchers and extension specialists, the app creates space to share ongoing work and highlight research gaps. Because instruments can be customized, results viewed immediately, and data exported for deeper analysis, the information gathered can feed directly into local, regional, or national strategies. The app makes it easier for researchers from different backgrounds to bring their perspectives together and develop shared, practical steps for understanding and mitigating emerging and existing threats to plant health.

## 4 | DISCUSSION

The MetaQuestion app gives organizers, policy makers, and stakeholders a straightforward way to build instruments that focus on the important issues in plant disease ecology and management. The systematic integration of experiential knowledge and expertise through expert knowledge elicitation captures nuanced, on-the-ground knowledge often not available from other resources. Extension and outreach personnel are pivotal sources for the acquisition and dissemination of such non-codified knowledge. It is also important to use appropriate caution in interpreting the results of expert knowledge elicitation, due to the often high uncertainty associated with expert opinions (Morgan 2014; O'Hagan 2019; Robledo et al. 2025). Formalizing expert knowledge elicitation supports a more holistic and adaptive knowledge system for plant health and applied plant ecology.

MetaQuestion, applied using the PlantQuest question catalog, provides a flexible platform for expert knowledge elicitation in plant health, addressing key challenges in this field: converting expert knowledge into structured and analyzable data while supporting transparency and reproducibility (Robledo et al 2025). Although online platforms such as Qualtrics are widely used, they require adaptation to efficiently gather expert inputs on pest and pathogen ecology, pest management options, geographic context, and networks of dispersal or trade. MetaQuestion streamlines expert knowledge elicitation by i) incorporating a curated question catalog, ii) enabling easy creation and export of specialized question types, and iii) providing functionality to work offline. These components make the application of expert knowledge elicitation in applied plant ecology more efficient and reproducible.

The curated catalog for plant health, PlantQuest, is designed to reduce common pitfalls in expert knowledge elicitation such as ambiguous wording, inconsistent scales, and underspecified context (PlantQuest Team, in preparation). Higher-quality questions are a resource that can be reused, adapted, and expanded across projects and disciplines, enabling organizers to more quickly generate instruments while reducing redundancy, improving comparability across studies, and supporting reproducibility. Rapid analysis can support rapid responses, such as in the context of disaster plant pathology, when decisions about how to protect plant health must be made quickly (Etherton et al. 2024; Mouafo-Tchinda et al. 2025). Organizers of expert knowledge elicitation can propose new questions for inclusion in the PlantQuest catalog, for consideration by the PlantQuest Team. This allows for ongoing development and improvement of the PlantQuest catalog, with careful human review (Fig. 1). There is also the potential to incorporate large language models in the process of improving question catalogs, to accelerate the process while audited by the PlantQuest Team.

MetaQuestion integrates lessons learned from past expert knowledge elicitation efforts in plant health, particularly in generating complex question types and producing ready to use datasets for analysis (Robledo et al. 2025). Data cleaning is often a bottleneck in platforms like Google Forms and Qualtrics. MetaQuestion streamlines this step by directly generating structured outputs, minimizing errors, and shortening the pipeline to results. For example, MetaQuestion facilitates creation of map-based adjacency question types, producing adjacency matrices formatted for direct use in common software such as R and Python.

In environments where connectivity is sporadic, use of traditional internet-based tools that depend on stable connectivity is problematic. MetaQuestion addresses this challenge through its offline functionality. Experts can pre-download the instrument, complete questions offline, and later upload their responses. This feature is particularly relevant in global plant health contexts, where researchers often rely on contributions from experts working in resource-limited or isolated areas (Robledo et al 2025).

MetaQuestion supports interoperability. Surveys can be exported to Qualtrics and other platforms, allowing users to combine MetaQuestion's features with the established infrastructure of widely adopted tools. This flexibility facilitates integration into existing workflows. Together, these developments position MetaQuestion as part of a broader effort to integrate expert knowledge more systematically into plant health decision making (Andersen Onofre et al. 2021). Applications include parameterizing epidemiological models, informing risk assessments, and supporting scenario analysis (Wiebe et al. 2018) with tools in the R2M toolbox (www.garrettlab.com/r2m) such as impact network analysis (Etherton et al. 2023; Etherton et al. 2025; Garrett 2021).

Ongoing development of MetaQuestion will include adding more map standardization for map-type questions and more data visualization methods for summarizing responses. Addressing measures of expert confidence and uncertainty more directly will be an important addition. There is also the potential for incorporating other curated question catalogs in MetaQuestion.

MetaQuestion targets crucial bottlenecks in expert knowledge elicitation and enhances decision making by providing a range of question formats, drawing on a curated question catalog such as PlantQuest, and providing offline access and interoperability with other tools. **Note that the version of MetaQuestion associated with this preprint version incorporates a preliminary version of the PlantQuest question catalog; a version of MetaQuestion available in the near future, announced in an update to this preprint, will incorporate the upcoming PlantQuest version 1.0.**

## AUTHOR CONTRIBUTIONS

Robert Fontan developed the current React web application. Christopher Perez developed a prototype of the app in R. Ashish Adhikari, Romaric Mouafo-Tchinda, Aaron Plex Sulá, Jacobo Robledo, and Karen Garrett have led the PlantQuest Team, providing preliminary draft question formats. All authors provided design ideas for the app. All authors contributed to paper writing.

## ACKNOWLEDGEMENTS

We appreciate support from USDA Animal and Plant Health Inspection Service (APHIS) Cooperative Agreements AP21PPQS&T00C195 and AP22PPQS&T00C133 and the USDA Specialty Crop Research Initiative, project award nos. 2020-51181-32198, 2022-51181-38242 and 2024-51181-43302, from the U.S. Department of Agriculture's National Institute of Food and Agriculture. We also appreciate support from the CGIAR Seed Equal Research Initiative, USAID Bureau of Humanitarian Assistance (BHA) award number 720BHA22IO00136, and the CGIAR Trust Fund (www.cgiar.org/funders/); we thank all donors and organizations which globally support the work of CGIAR through their contributions to the CGIAR Trust Fund. The opinions expressed in this article are those of the authors and do not necessarily reflect the views of USAID or USDA. We appreciate helpful comments from T. Johnson, J. Malosh, A. Reddy Banda, S. Rickards, and I. Zerry.

## CONFLICT OF INTEREST STATEMENT

The authors declare no conflicts of interest.

# DATA AVAILABILITY STATEMENT

MetaQuestion is available at: https://r2meke.netlify.app/
**Note that the version of MetaQuestion associated with this preprint version incorporates a preliminary version of the PlantQuest question catalog; a version of MetaQuestion available in the near future, announced in an update to this preprint, will incorporate the upcoming PlantQuest version 1.0.**

Code will be available soon in an open repository link:
https://github.com/GarrettLab/EKE

# ORCID

*Ashish Adhikari* https://orcid.org/0000-0003-1016-0617
*Romaric A. Mouafo-Tchinda* https://orcid.org/0000-0003-1343-4820
*Aaron I. Plex Sulá* https://orcid.org/0000-0001-7317-3090
*Jacobo Robledo* https://orcid.org/0000-0002-2707-4005
*Karen A. Garrett* https://orcid.org/0000-0002-6578-1616

# REFERENCES

Andersen Onofre, K. F., Forbes, G. A., Andrade-Piedra, J. L., Buddenhagen, C. E., Fulton, J., Gatto, M., Khidesheli, Z., Mdivani, R., Xing, Y., and Garrett, K. A. 2021. An integrated seed health strategy and phytosanitary risk assessment: Potato in the Republic of Georgia. Agric. Syst. 191:103144.

Courtney Jones, S. K., Geange, S. R., Hanea, A., Camac, J., Hemming, V., Doobov, B., Leigh, A., and Nicotra, A. B. 2023. IDEAcology: An interface to streamline and facilitate efficient, rigorous expert elicitation in ecology. Methods in Ecology and Evolution 14:2019-2028.

Dicks, L. V., Breeze, T. D., Ngo, H. T., Senapathi, D., An, J., Aizen, M. A., Basu, P., Buchori, D., Galetto, L., and Garibaldi, L. A. 2021. A global-scale expert assessment of drivers and risks associated with pollinator decline. Nature Ecology & Evolution 5:1453-1461.

Etherton, B. A., Plex Sulá, A. I., Mouafo-Tchinda, R. A., Kakuhenzire, R., Kassaye, H. A., Asfaw, F., Kosmakos, V. S., McCoy, R. W., Xing, Y., Yao, J., Sharma, K., and Garrett, K. A. 2025. Translating Ethiopian potato seed networks: identifying strategic intervention points for managing bacterial wilt and other diseases. Agric. Syst. 222:104167.

Etherton, B. A., Choudhury, R. A., Alcalá-Briseño, R. I., Xing, Y., Plex Sulá, A. I., Carrillo, D., Wasielewski, J., Stelinski, L., Grogan, K. A., Ballen, F., Blare, T., Crane, J., and Garrett, K. A. 2023. Are avocados toast? A framework to analyze decision-making for emerging epidemics, applied to laurel wilt. Agric. Syst. 206:103615.

Etherton, B. A., Choudhury, R. A., Alcalá Briseño, R. I., Mouafo-Tchinda, R. A., Plex Sulá, A. I., Choudhury, M., Adhikari, A., Lei, S. L., Kraisitudomsook, N., Robledo

Buritica, J., Cerbaro, V. A., Ogero, K., Cox, C. M., Walsh, S. P., Andrade-Piedra, J., Omondi, B. A., Navarrete, I., McEwan, M. A., and Garrett, K. A. 2024. Disaster plant pathology: Smart solutions for threats to global plant health from natural and human-driven disasters. Phytopathology 114:855-868.

Garrett, K. A. 2021. Impact network analysis and the INA R package: Decision support for regional management interventions. Methods in Ecology and Evolution 12:1634-1647.

Hemming, V., Burgman, M. A., Hanea, A. M., McBride, M. F., and Wintle, B. C. 2018. A practical guide to structured expert elicitation using the IDEA protocol. Methods in Ecology and Evolution 9:169-180.

Martin, T. G., Burgman, M. A., Fidler, F., Kuhnert, P. M., Low-Choy, S., McBride, M., and Mengersen, K. 2012. Eliciting expert knowledge in conservation science. Conservation Biology 26:29-38.

McBride, M. F., Garnett, S. T., Szabo, J. K., Burbidge, A. H., Butchart, S. H., Christidis, L., Dutson, G., Ford, H. A., Loyn, R. H., and Watson, D. M. 2012. Structured elicitation of expert judgments for threatened species assessment: a case study on a continental scale using email. Methods in Ecology and Evolution 3:906-920.

Morgan, M. G. 2014. Use (and abuse) of expert elicitation in support of decision making for public policy. Proceedings of the National Academy of Sciences 111:7176-7184.

Mouafo-Tchinda, R., Etherton, B., Plex Sula, A., Andrade-Piedra, J., Ogero, K., Omondi, B. A., McEwan, M., Tene Tayo, P., Harahagazwe, D., Cherinet, M., Gebeyehu, S., Sperling, L., and Garrett, K. A. 2025. Pathogen and pest risks to vegetatively propagated crops in humanitarian contexts: Toward a national plant health risk analysis for Cameroon and Ethiopia. bioRxiv https://doi.org/10.1101/2024.02.12.580019

O'Hagan, A. 2019. Expert knowledge elicitation: subjective but scientific. The American Statistician 73:69-81.

PlantQuest Team. in preparation. PlantQuest v1.0: A curated question catalog for expert knowledge elicitation addressing plant health and applied plant ecology.

R Core Team. 2025. R: A language and environment for statistical computing. R Foundation for Statistical Computing, Vienna, Austria.

Robledo, J., Plex Sulá, A. I., Jaworski, L. G., Mouafo-Tchinda, R. A., Andersen Onofre, K. F., Thomas-Sharma, S., and Garrett, K. A. 2025. Expert knowledge elicitation: Accessing the big data in experts' brains. Phytopathology 115: https://doi.org/10.1094/PHYTO-06-25-0220-FI.

Wiebe, K., Zurek, M., Lord, S., Brzezina, N., Gabrielyan, G., Libertini, J., Loch, A., Thapa-Parajuli, R., Vervoort, J., and Westhoek, H. 2018. Scenario development and foresight analysis: exploring options to inform choices. Annual Review of Environment and Resources 43:545-570.